\documentclass[aps,pra,10pt,english,showpacs,floatfix,twocolumn,twoside]{revtex4-1}

\usepackage{amsmath,amstext,amssymb,babel,graphicx,geometry,bm,dcolumn,color,hyperref}
\usepackage{hyperref}
\hypersetup{colorlinks=true, urlcolor=blue}
\usepackage{mathtools}

\def \be{\begin{equation}}
\def \ee{ \end{equation} }

\usepackage{epstopdf}

\begin{document}
\renewcommand*{\DefineNamedColor}[4]{%
  \textcolor[named]{#2}{\rule{7mm}{7mm}}\quad
  \texttt{#2}\strut\\}

\definecolor{red}{rgb}{1,0,0}
\title{Benford analysis of quantum critical phenomena: First digit provides high finite-size scaling exponent while first two and further are not much better}

\author{Anindita Bera\(^{1,2}\), Utkarsh Mishra\(^{3}\), Sudipto Singha Roy\(^{2}\),\\ Anindya Biswas\(^{4}\), Aditi Sen(De)\(^{2}\) and Ujjwal Sen\(^{2}\)}

\affiliation{\(^1\)Department of Applied Mathematics, University of Calcutta, 92 Acharya Prafulla Chandra Road, Kolkata 700 009, India\\
\(^2\)Harish-Chandra Research Institute, HBNI,  Chhatnag Road, Jhunsi, Allahabad 211 019, India\\
\(^3\)Asia Pacific Center for Theoretical Physics, Pohang, Gyeongbuk 790-784, Korea\\
\(^4\)Department of Physics, National Institute of Technology Sikkim, India}

\begin{abstract}
Benford's law is an empirical edict stating that the lower digits appear
more often than  higher ones as the first few significant digits in statistics
of natural phenomena and mathematical tables. 
A marked proportion of such analyses is restricted to 
the first significant digit.
We employ violation of Benford's law, up to the first four significant
digits, for investigating
magnetization and correlation data of paradigmatic quantum many-body
systems to detect cooperative phenomena, focusing on the finite-size
scaling exponents thereof. 
We find that for the transverse field quantum \(XY\) model, behavior of the very first significant digit of 
an observable, at an arbitrary point of the parameter space, is enough 
to capture the quantum phase transition in the model with a relatively high
scaling exponent. A higher number of significant digits do not
provide an appreciable further advantage, in particular, in terms of an increase in
scaling exponents. Since the first significant digit of a physical quantity is relatively 
simple 
to obtain in experiments, the results have potential implications for laboratory observations
 in noisy environments.
\end{abstract}

\maketitle

\section{Introduction}
\label{sec:benfors_law}
The leading digit phenomenon, discovered by  Newcomb in 1881~\cite{newcomb} and then independently
by Benford in 1938~\cite{benford}, states that the frequencies of occurrence of the first significant digits, in many  natural and mathematical  sets of data points are not random, and instead  follows a pattern due to which the smaller digits occur more often than the  larger ones. This statement is contrary to the common intuition that each of the digits from $1$ to $9$ has an equal probability 
(i.e. about \(11.1\%\)) of being the first significant digit in a number.  Newcomb and Benford, after analyzing several data sets, found that 
the digit $1$ has about $30\%$ chance to occur  as  the first significant digit.
The percentages monotonically decrease for larger digits, and for example the digit $9$ was observed to occur with a probability less than $5\%$. 
The observations can be cast in terms of a probability distribution function. The probability of occurrence of the digit $d$ as the first significant digit can be expressed, according to Benford's law, as
\begin{eqnarray}
\label{eqn:fsdl}
P(D_1)=\log_{10}(1+1/D_{1}), \quad  D_{1}  \in \{ 1,2,\ldots,9\}.
\end{eqnarray}
Within  this formula, $P(D_1=1)=\log_{10}(2) = 0.3010 \ldots,$
$P(D_1=2)=\log_{10}\frac{3}{2}=0.1760 \ldots,$ and so on. At the end of the spectrum, $P(D_1=8)=\log_{10}\frac{9}{8} = 0.05115 \ldots$ and  
$P(D_1=9)=\log_{10}\frac{10}{9}=0.04575\dots$. This shows that the two smallest digits, 
$1$ and $2$, occur with a combined probability close to $50\%$, whereas the two largest digits
together have a probability  less than $10\%$.

Benford's law,  given by Eq.~(\ref{eqn:fsdl}), is satisfied by data from a large number of sources. 
The Benford probability distribution has been checked 
for various physical constants~\cite{Knuth, BurkeandKincanon},
half-lives of 477 radioactive substances~\cite{Bucketal}, financial and 
accounting data~\cite{account,account1,account2}, etc.  Moreover, the  law has also been tested for  numbers 
drawn from common mathematical distributions and observed to be valid 
for several.   
Interestingly, several sequences, e.g., the Fibonacci sequence, have been found to follow 
Benford's law~\cite{fibonacci}. An emergent connecting feature of data satisfying Benford's law seems to be that such data must contain numbers of several orders of magnitude. 

Notwithstanding the empirical validity of the law for data obtained from a variety of sources, there are several examples of data sets where Benford's law gets violated.
Violation of Benford's law  is useful  in detecting frauds in 
data manipulation cases, for example, in tax 
payers' returns~\cite{account1}, elections~\cite{account2}, etc. See \cite{Nigrini-thesis} in this respect. It has been observed that un-manipulated  statistical data follow the Benford prediction, while manipulated data  deviate from same. Apart from this, violation of Benford's law can be used to detect faint earthquakes, which was explored by Sambridge
\emph{et al.}~\cite{earthquake}. In this case, the background noise tends to violate the Benford predictions,  while the seismometer readings during   the actual earthquake tend to follow the same.
Over the years, many theoretical studies have been carried out  to unveil fundamental characteristics of a  distribution which satisfies  Benford's prediction~\cite{Pinkham61, hill_95, benford_cond1, Torres2007,Formann2010}. In particular, in Ref.~\cite{benford_cond1}, it was shown that Benford's law can also be satisfied for data generated from deterministic dynamics. 
A rather complete list of papers dealing with the Benford's law can be found at \cite{Hillwebsite}.

A more complete form of the Benford's probability law, expressed in Eq.~(\ref{eqn:fsdl}), 
can be given in terms of the joint distribution of  the first $k$ significant digits~\cite{hill_95}.
In general,  let $D_{1}, D_{2}, \ldots, D_{k}$ denote the first, second, $\dots$, $k^{\text{th}}$ significant digits 
of a number. Then the generalized Benford's law is given by
\begin{equation}
\label{eq:gsdl}
P(D_{1}=d_{1},\ldots,D_{k}=d_{k})=\log_{10}\big[1+\big(\sum_{i=1}^{k}d_{i}\times10^{k-i}\big)^{-1}\big]. 
\end{equation}
Here, $d_1$ is an integer that belongs to the numbers from the set $\{ 1,2,\ldots,9\}$, and for 
$i\geq 2$, $d_{i}$ belongs to set $\{ 0,1,2,\ldots,9\}$. Note  that the base of the
logarithm here is again $10$.  One can easily  see that for $k=1$, the 
above equation reduces to Eq.~(\ref{eqn:fsdl}). The probability distribution $P(D_{1}=d_{1},\ldots,D_{k}=d_{k})$ becomes more and more flat and
uniform as the number of digits, i.e., $k$, increases. At this stage, the notable fact is  that the knowledge of these higher significant digits indeed provides more information about a 
number, as compared to just the first significant digit. We will however show that such information may not be substantial for Beford analysis of certain physical phenomena. From this expression, one can also deduce corresponding probabilities of the second, third, and higher digits being one of the digits from the set $\{0,1,\ldots, 9\}$, by summing over the other random variables   in the expression. However, it seems that knowing the probability of occurrence of  a particular combination of digits, say  $(D_1,D_2,D_3)$,  is more informative in comparison to the situation when the information of occurrence of only a portion of the combination is known. For instance, 
consider a number $x$, written in decimal representation as  $x=0.d_1d_2\dots \times 10^n$, where $d_1 \in \{1,2,\dots,9\}$, $d_i \in \{0,1,2,\dots,9\}$, for $2 \leq i \leq l$, with $n$ and $l$ being integers.
If one does not know the first significant digit, say $d_1$, then by knowing $d_2$ or having information about  any other single digit $d_{i},~2 \leq i \leq l$, does not provide much information 
about the number as compared to the case when more that one digits $d_{j},~j\geq1$, are known. Therefore, instead of  comparing the Benford distribution of the $k^{\text{th}}$ significant digit, we compare the Benford distribution of the first $k$ significant digits.

 The Benford's 
law of first significant digit, expressed in Eq.~(\ref{eqn:fsdl}), has been found to be useful in detecting
the quantum phase transition (QPT) points of  one-dimensional  quantum Ising  and anisotropic quantum  $XY$ models, in transverse magnetic field~\cite{au_benford,utkarsh_pre}. A drastic change in the observed distribution of the first significant digit of physical observables like transverse magnetization, classical correlation functions and entanglement  are noticed near the QPT~\cite{au_benford}.
The transverse magnetization data e.g., show a violation of Benford's law for the entire 
range of the one-dimensional parameter space, and the violation amounts are relatively stable with respect to the variation of the parameter except near the critical point (QPT point).
The violation amounts are different on 
two sides of the QPT, and an abrupt (single-period) oscillatory behavior characterizes the critical 
point. 
A similar feature was noticed for Benford analysis of earthquake data in Ref.~\cite{earthquake}. The findings have an interesting significance. It is to be noted that the transverse magnetization of the ground state of the one-dimensional transverse field quantum $XY$ model can detect the quantum phase transition in the model. However, unlike the (parallel) magnetization of the symmetry-broken ground state, the transverse megnetization does not change from being zero to non-zero at the QPT. On the other hand, the Benford violation parameter of the transverse  magnetization has a behavior, near the QPT, that is quite close to the behavior of the parallel magnetization, in that the former saturates to two different values on two sides of the QPT. A finite-size scaling of the violation parameter of transverse magnetization was performed in Ref.~\cite{utkarsh_pre} and it was obtained that analysis of the first digit already provides a comparatively high value of the finite-size s
 caling exponent for the phase transition in the model.

In this work,  
we raise the following question:
\emph{Does  going beyond the first significant digit provide any further increment in the scaling exponent?}

To this end, we focus on the generalized Benford's law given in Eq.~(\ref{eq:gsdl}), and investigate the quantum phase transition of the one-dimensional anisotropic quantum  $XY$ model in transverse magnetic field, using the knowledge of the first few
significant digits of transverse magnetization data and its Benford analysis. We perform  Benford analysis of the data obtained from the variation of the  transverse magnetization, 
up to first four significant digits.  We find that the  scaling exponent obtained from the first significant digit is optimal, in the sense that no further advantage is obtained.
The results therefore have the ``positive'' significance to an actual observer in the sense that it reduces her/his measurement task. 

It is important to present here a comparison of the results obtained in this paper with those in Refs.~\cite{au_benford,utkarsh_pre}. In Ref.~\cite{au_benford}, it was 
reported that Benford analysis of the first significant digit of physical observables like magnetization, classical correlations, and entanglement is capable of detecting quantum phase transition (QPT) in some  
strongly correlated systems. The feature obtained is very similar that in the Benford analysis of  earthquake data.  Moreover, it was also noticed that the behaviors of the relative frequencies of the first significant digits
for the transverse magnetization of the infinite transverse Ising model are significantly different. Subsequently, in Ref.~\cite{utkarsh_pre}, a finite-size scaling analysis of the Benford violation parameter of transverse magnetization 
was performed and it was obtained that an analysis of the first digit provides a comparatively high value of the finite-size scaling exponent near the phase transition point in the model. The significance of this finding lies in the fact 
that finite-size scaling analysis provides an impression  about the closeness of a finite system to its thermodynamic limit, and a high scaling exponent implies that one can mimic the infinite size system even at relatively small system-size. 
However, the question which remained unanswered in the two works is the following. If instead of the first significant digit, one considers the first two, first three, first four, $\ldots$ higher significant digits, for detection of the 
quantum phase transition, would there be any improvement in the scaling exponent? This is an important question because discerning the first significant digit of an observed quantity is the easiest in an experiment. The first two is less 
easy, and so on. Therefore, it is important to know for an experimentalist as to how many significant digits need to be observed for the physical quantity under study, to pin down the point of phase transition. 
The answer to this question cannot be obtained or implied from the results in Refs.~\cite{au_benford,utkarsh_pre}, and indeed new sets of data are needed for the corresponding analyses. 

The paper is organized as follows. In Sec.~\ref{sec:model_settings}, we provide details of
the anisotropic quantum $XY$ model and the corresponding  expression of  transverse magnetization and other related quantities.
A short description of the Benford methodology is presented in Sec.~\ref{sec:benford_setting}. 
For completeness, in Sec.~\ref{sec:first-digits},  we discuss the application of  Benford's law  for the  first significant digit in detection of QPT in the transverse field anisotropic $XY$ model.
Sec.~\ref{sec:higher-digits} reports the patterns of violation of the  
generalized Benford's law for up to four significant digits 
in the $XY$ model, and present a  study of behavior of the scaling exponent with respect to increase 
of significant digits in the data set.
We present a conclusion in Sec.~\ref{sec:conc}.

\section{The model}
\label{sec:model_settings}
In this section, we start with a brief introduction of the model Hamiltonian that we have considered for our work. The Hamiltonian of the anisotropic transverse quantum  $XY$ model in a one-dimensional (1D) lattice reads~\cite{LSM,BMD, amit_dutta} 

\begin{equation}
 H = \dfrac{J}{4} \displaystyle \sum_{i=1}^N \left[(1+\gamma)\sigma_{i}^{x}\sigma_{i+1}^{x}
 +(1-\gamma)\sigma_{i}^{y}\sigma_{i+1}^{y}\right] - \frac{h}{2} \sum_{i=1}^N \sigma_{i}^{z},
 \label{eq:ising_H}
\end{equation}
where $J$ is proportional to the coupling constant, 
$h$ is proportional to the strength of the transverse magnetic field, $\gamma~(\ne0)$ is the anisotropy parameter, $\sigma$'s are the Pauli matrices, and  $N$ is the number of sites in the lattice.
We assume periodic boundary condition, i.e, $\sigma_{N+1}=\sigma_{1}$. The anisotropic quantum
$XY$ model for $\gamma \ne 0$ forms the ``Ising universality class". 
For $\gamma=1$, the Hamiltonian, described by Eq.~(\ref{eq:ising_H}), is known as the transverse
Ising Hamiltonian. The model is diagonalizable by applying successive Jordan-Wigner, Fourier, and 
Bogoliubov transformations~\cite{LSM,BMD,BM}. It is known that the system undergoes a quantum phase transition at $\lambda=\lambda_{c}\equiv 1$~\cite{LSM, BMD,BM} from  a long-range antiferromagnetic ($\lambda<1$) to a paramagnetic phase ($\lambda >1$), where we have assumed that $J>0$.

The quantum critical point of the model is 
detected by looking at the behavior
of certain physical properties of the zero-temperature state in the thermodynamic limit. 
For instance, the paramagnetic to antiferromagnetic transition has been 
detected using two-site entanglement (one among many methods) 
and it is found that the derivative of the entanglement, 
with respect to the driving parameter $\lambda$, diverges at the critical point, signaling the 
presence of a QPT~\cite{osterloetal, Nielsonetal}.
A single-site observable like transverse magnetization can also be used to detect this QPT by
computing the derivative with respect to the driving parameter $\lambda$ as well as by 
analyzing the data using the Benford technique. 
For finite spin systems,
the transverse magnetization is given by
 \begin{equation}
	M_{z}(\lambda,\widetilde{\beta},N)=-\frac{2}{N}\sum_{p=1}^{N/2}\frac{\tanh(\widetilde{\beta}\Lambda_p(\lambda)/2)
	(\cos\phi_{p}-\lambda)}{\Lambda_p(\lambda)},
	\label{eq:mz_finite}
\end{equation}
where  $\widetilde{\beta}=\beta J$ with $\beta=\frac{1}{k_B T}$, $k_B$ being the Boltzmann constant, $T$ being the absolute temperature, 
 $\phi_{p}=\frac{2\pi p}{N}$,
and $\Lambda_p (x)=\left\{\gamma^2\sin^2\phi_p+[x-\cos\phi_p]^{2}\right\}^{1/2}$.
In the thermodynamics limit, the transverse magnetization is given by
\begin{equation}
\label{eq:fintemp}
	M_{z}(\lambda,\widetilde{\beta})=-\frac{1}{\pi}\int\limits_{0}^{\pi}d\phi\frac{\tanh(\widetilde{\beta}\Lambda
	(\lambda)/2)(\cos\phi-\lambda)}{\Lambda(\lambda)},
\end{equation}
where  
\begin{equation}
\Lambda (x)=\left\{\gamma^2\sin^2 \phi+[x-\cos\phi]^{2}\right\}^{1/2}.
\end{equation}
The two-site spin-spin correlation functions can also be calculated analytically for this model for both finite and
infinite lattice sizes at any temperature. The nearest-neighbor diagonal correlation functions are given by
 \begin{eqnarray}
   T_{xx}(\lambda)=G(-1,\lambda),~T_{yy}(\lambda)=G(1,\lambda),
 \label{eq:cxx}
\end{eqnarray} and
 \begin{equation}
 	T_{zz}(\lambda)=[M_{z}(\lambda)]^{2}-G(-1,\lambda)G(1,\lambda),	
 \end{equation}
 where $G(R,\lambda)$, at zero temperature and $N \to \infty$, is given by  
 \begin{equation}
 	G(R,\lambda)=\frac{1}{\pi}\int\limits_{0}^{\pi}d\phi\frac{(\gamma\sin(\phi R)\sin\phi-\cos(\phi R)(\cos\phi-\lambda))}{\Lambda(\lambda)}.
 	\label{eq:GR}
 \end{equation}
The magnetization, two-site spin-spin correlation functions, and bipartite entanglement are smooth functions of the driving parameter $\lambda$. However, the derivatives of the above quantities with respect to $\lambda$ show sharp kinks at the critical point.  If a finite size scaling of the above quantities is performed, it is observed that the critical point is reached with the scaling law 
\begin{equation}
\lambda_{c}^N\approx \lambda_{c}+\alpha N^{-q},
\end{equation}
where $\alpha$ is a constant. The scaling  exponent $q$  governs the thermodynamic properties in the sense that its value indicates the pace with which the system approaches to its thermodynamic critical point with increasing  system size. 
In the succeeding section, we discuss the Benford technique in order to detect  the quantum critical point and corresponding values of the scaling exponent, $q$, by choosing transverse magnetization and correlations as the observables. 

\section{Benford analysis of physical data}
\label{sec:benford_setting}
In this section, we outline the method to analyze physical data using the Benford's law.  
First we focus on the transverse magnetization, $M_{z}(\lambda)$,  at zero temperature which is given by
Eq.~({\ref{eq:fintemp}}), for a range of interest, say $[a,b]$  of the parameter $\lambda$. The data could also be obtained from an actual experiment. We now consider a shifting magnetic field window of length $w$ inside the interval $[a,b]$. The generic form of the window is 
\begin{equation}
[a+m\epsilon,a+w+m\epsilon],
\end{equation}
where $m$ runs over non-negative integers  until $a+w+m\delta=b$. Typically, we consider $0 < \epsilon < w \ll b-a$. Note therefore that consecutive windows, for consecutive values of $m$, overlap to a significant extent. Compare the shifting field window here with the shifting time window for Benford analysis of seismic data in Ref.~\cite{earthquake}. We are interested in the first significant digit of the 
magnetization data in each subinterval. 
Since the subinterval length, $w$, is very small in comparison with the total interval $[a,b]$, it is unlikely for the first significant digit to vary  a lot within this range. As a consequence, an analysis of such data would result in a trivial violation of Benford's law with uninteresting implications.
Therefore, we  normalize the data for each subinterval in such a way that 
the value of the physical quantity lies between $0$ and $1$. The normalized data in a given  subinterval is given by

\begin{eqnarray}
M^{B}_{z}(\lambda)=\frac{M_{z}(\lambda)-M_{z}^{min}(\lambda)}{M_{z}^{max}(\lambda)-M_{z}^{min}(\lambda)}, 
\end{eqnarray}
 where $M_z^{min}(\lambda)$ and $M_z^{max}(\lambda)$ are respectively the minimum and maximum magnetization in the chosen subinterval.
The magnetization, $M^{B}_{z}(\lambda)$, obtained in this way is termed as 
``Benford magnetization''~\cite{au_benford,utkarsh_pre}.
Once we obtain the Benford magnetization, we are equipped with the tools to analyze the data using the frequency of  occurrence of the first  significant digit. For the chosen subinterval,  let us suppose that  the  observed frequency of the digit $D_1$ as the first significant digit is  $O_{D_1}$ and the expected frequency of the digit $D_1$ as first significant digit is $E_{D_1}$.  We now wish to estimate the degree to which the observed data violates Benford's law. One way do so, which is reminiscent of the concept of mean deviation, is to consider the quantity
\begin{equation}
\Delta_{md}(M^{B}_{z})=\sum_{D_1}\Big|\frac{O_{D_1}-E_{D_1}}{E_{D_1}}\Big|,
\label{eq:bvp}
\end{equation}
where we remember that the entire analysis is being performed for a given subinterval $[a+m\epsilon,a+w+m\epsilon]$, i.e., for a given $m$. We consider $\Delta_{md}(M_z^B)$ to be a function of the midpoint of the subinterval, i.e. of $\lambda=a+\frac{w}{2}+m\epsilon$. The suffix ``$md$'' is to remind us that the distance between the observed data and Benford's prediction, as quantified by $\Delta_{md}$, is akin to the mean deviation.  If there are $n$ observed data points in each
subinterval, then the expected distribution for the significant digit 
$D_1$ is $E_{D_1} = n\log(1+1/D_1)$. The analysis can be generalized to the first two, first three, 
\(\ldots\) significant digits. We refer to the quantity $\Delta_{md}(M^B_{z})$ as a  Benford violation parameter (BVP) for  magnetization~\cite{au_benford, utkarsh_pre}. Notice that we use the 
same symbol \(M_z^B\) for Benford transverse magnetization corresponding to any number of 
first significant digits, to keep the notation uncluttered. The number of significant digits in a given situation will always be made clear by the context. [A similar approach for notation is 
followed below for the finite-size scaling exponents.]
Note also that below we will introduce and consider other parameters (measures) of Benford violation also. We now vary $m$ to obtain the violation parameter for other magnetic field windows, and in this way, we may obtain the profile of the violation parameter for the entire magnetic field range $[a,b]$. For a given range $[a,b]$, and for given $w$ and $\epsilon$, we need to check for convergence of the violation parameter profile with respect to $n$. As mentioned before, 
it is observed that the frequency distribution of the significant digits approaches to a uniform distribution as $k$, the number of significant digits, is increased~\cite{hill_95b}. Therefore, it is often enough to consider Benford violation in a given data set for up to first four significant digits, as  beyond that  the distribution is more likely to follow a uniform distribution. We therefore restrict our analysis   to the first four significant digits. 
A similar procedure can also be followed to find the Benford violation parameter for other 
observables of the system. For example, we also 
calculate spin-spin correlators, $T_{xx}^{j,j+1}$, between nearest-neighbor sites and the corresponding
Benford violation parameter, $\Delta_{md}(T^{B}_{xx})$.


In this paper, in addition to the mean deviation-based distance, we also use  distances inspired by the standard deviation
and the Bhattacharyya distance~\cite{bhattacharya}.
The standard deviation-based distance  between the frequency distributions $O_{D_1}$ and $E_{D_1}$ reads

\begin{equation}
\Delta_{sd}=\sqrt{\sum_{D_1=1}^9(O_{D_1}-E_{D_1})^2}.
\label{eq:sd}
\end{equation}
On the other hand,  the Bhattacharyya distance has following mathematical expression:
\begin{equation}
\Delta_{bd}=-\ln\sum_{D_1=1}^9 \sqrt{O_{D_1} E_{D_1}}.
\label{eq:bd}
\end{equation}

\section{Benford violation for first significant digit in the $XY$ model}
\label{sec:first-digits}
In this section, for completeness, we  review important results obtained previously from the  Benford analysis of the first significant digit of magnetization of the transverse $XY$ model.
In general, the onset of a QPT is indicated by certain changes in the behavior of the zero-temperature state of a many-body system when an external parameter of the Hamiltonian is varied~\cite{subir_sachdev}. In recent years,  Benford's law of leading digits has also been used to detect such phase transitions in the quantum $XY$ model using the data of two-site entanglement and single-site magnetization~\cite{utkarsh_pre,au_benford}. The  Benford violation parameter corresponding to transverse magnetization, $M_z$, with respect to the driving parameter $\lambda$  exhibits an abrupt change at the  critical point, $\lambda_c$~\cite{au_benford, utkarsh_pre}. A similar behavior
was observed using quantum entanglement~\cite{au_benford}.  The finite-size scaling exponent, $q$, turns out to be comparatively large when it is obtained from Benford analysis of the first significant digit of $M_z$, in comparison to the exponent obtained from  the derivative of $M_z$. In particular,  the value of the finite-size scaling  exponent obtained  from the Benford transverse magnetization turns out to be $2.06$ for $\gamma=0.5$, while the same obtained from  scaling behavior of the derivative of $M_z$ with respect to the parameter $\lambda$ yields $q=1.67$. 
This shows the importance of Benford analysis of leading digits in capturing  quantum criticality in 
complex many-body systems.
\section{Benford violation beyond first significant digit in the $XY$ model}
\label{sec:higher-digits}
In the previous sections, we have summarized the results obtained from Benford analysis carried out for the first significant digit of physical observables and realized that BVP can serve as an efficient detector of  QPT points. In this section, we move one step further and raise the following question:
\emph{Is  it  worthwhile  to  go beyond the first significant digit?}

\begin{figure}
\includegraphics[angle=360,width=4.3cm,height=3.8cm]{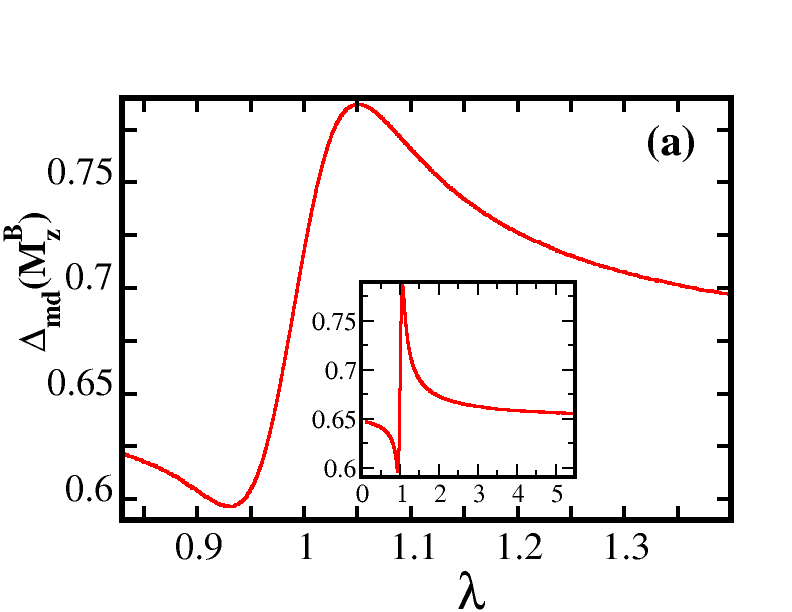}
\includegraphics[angle=360,width=4.3cm,height=3.8cm]{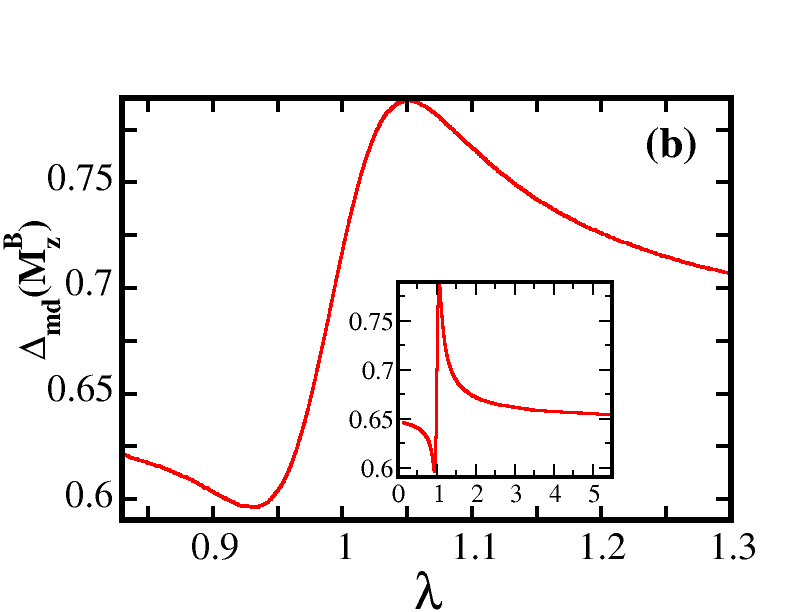}
\includegraphics[angle=360,width=4.3cm,height=3.8cm]{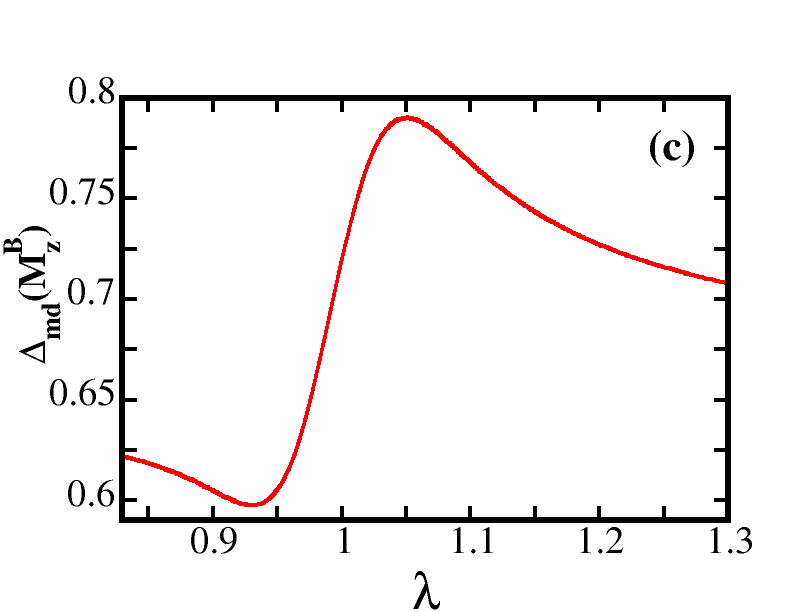}
\includegraphics[angle=360,width=4.3cm,height=3.8cm]{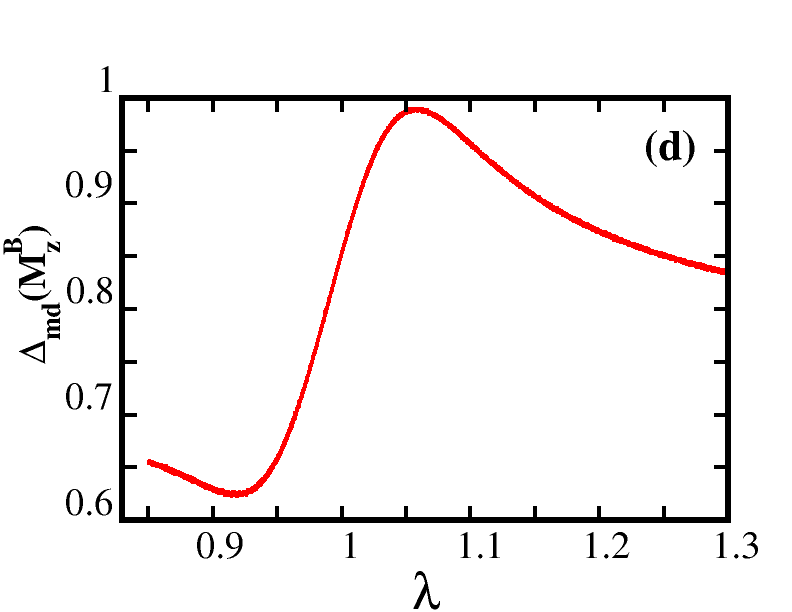}
\caption{(Color online.) Benford analysis of first few significant digits of transverse magnetization near a quantum phase transition in a spin model. We plot $\Delta_{md}(M^{B}_{z})$, the Benford violation parameter of the transverse magnetization, for different numbers of first few significant
digits, for the zero-temperature state of the transverse  quantum \(XY\) model. The panels have the Benford violation parameter on the vertical axes, while the horizontal axes represent \(\lambda\). The panels (a), (b), (c), and (d), respectively correspond to the first, first two, first three, and first four significant digits. The width of the field window is chosen to be $0.05$ in all cases. The data have been checked for convergence with respect to the number of points, \(n\), chosen within a given magnetic field window. For the curves displayed in the panels, 
\(n = 10^4, 10^4, 1.1 \times 10^4, 4 \times 10^4\) respectively. 
In all the cases, we have considered the system size \(N = 40\), and have chosen the anisotropy 
\(\gamma = 0.5\). 
The insets in panels (a) and (b) exhibit the behaviors of the functions in the 
corresponding main plots for a wider range of values of \(\lambda\).
All quantities are dimensionless.
}
\label{fig:fig1}
\end{figure}

To answer this, we  investigate the Benford violation parameter for $M_z$ obtained from the zero-temperature state of the anisotropic $XY$ model with a transverse magnetic field, by taking into account the first four significant digits of the magnetization data.  
The width of the field window is kept fixed at $w=0.05$. We choose $\epsilon=5\times 10^{-5}$. In addition to this, several values of $n$, the number of data points in a given subinterval of length $w$, are chosen to check for convergence of the Benford violation amount for that subinterval. 
We consider  systems of finite (periodic) chains consisting of $N$ spins.
The value of the anisotropy parameter is set at $\gamma=0.5$. 
Note here that the qualitative results remain invariant with the variation of $\gamma$.
The variation of the Benford violation parameter (BVP) for  different numbers of the first few significant digits of the transverse magnetization against the parameter 
$\lambda$ is depicted in Fig.~\ref{fig:fig1}. The BVP changes abruptly 
as the system 
crosses the quantum critical point at  $\lambda=\lambda_c \equiv 1$. The profile of the Benford violation parameter has a minimum just before the critical point and a maximum just after, and we checked that this is true irrespective of whether we consider the first, first-two, first-three, or first-four significant figures, for evaluating the violation parameter. This underlines the fact that the first significant digit already captures the most significant aspect of the transverse magnetization in the system with respect to the Benford violation  near the QPT. We will come back to this point when we consider the finite-size scaling exponents.\


We note here that in order to make the plots smooth, we need more data points within a small magnetic field window for computation of the converged BVP (convergence with respect to \(n\)) for higher numbers of digits, in comparison to what is required for the first digit. We plot the Benford violation parameter for the first four significant digits in Fig.~\ref{fig:fig1} (d). For this plot, we have considered $4\times 10^4$ data points in every field window of width $w=0.05$, while 
for the first significant digit, \(10^4\) data points are enough to 
obtain convergence. This is another advantage of using the first significant digit over the situation where more than one significant digits are considered. The plots shown in Fig.~\ref{fig:fig1} are obtained using the ``mean-deviation distance", $\Delta_{md}$, between the observed and Benford frequencies. The qualitative behavior of the Benford violation parameter, however, remains the same for other measures of distance, as given in Eqs.~(\ref{eq:sd}) and (\ref{eq:bd}).  It is observed that when the system is away from the critical point, the violation parameter for the above four cases remain almost constant with respect to the changing field window. If we consider a functional fit of the data, as shown in  Fig.~\ref{fig:fig1}, and calculate the derivative of BVP with $\lambda$, a sudden change in the system is signaled at $\lambda \approx 1$, which can be identified with the QPT 
in that system. 
For the first significant digit, this has been reported in~\cite{utkarsh_pre}.

For the 
finite-size scaling analysis,
we 
 first obtain the data for transverse magnetization and other relevant system characteristics, for different system sizes, viz. $N=14, 20, 24, 30, 34,$ and $ 40$, as functions of  $\lambda$.
For each $N$, the Benford violation 
parameter is calculated for every field window, as dictated by the window width and window shift chosen, and the results retained up to the first four
significant digits. 
\begin{center}%
\begin{table}[htb]
\begin{tabular}{|c|c|c|c|c|c|}
\hline
Significant & & & & \\ digits
& $q_{md}(	M_z)$ & $q_{sd}(M_z)$ & $q_{bd}(M_z)$ & $q_{md}(T_{xx})$ 
\\ \hline
\hline 
1    
&  2.06      &  2.30    &  2.55    &  2.04   \\ \hline
2    
&  2.10      &  2.31    &  2.72    &  2.07   \\ \hline
3    
&  2.11      &  2.30    &  2.72    &  2.07   \\ \hline
4    
&  2.09      &  2.5     &  2.8     &  2.08   \\ \hline
\end{tabular}
\caption{
Finite-size scaling exponents for the
transverse magnetization and classical correlation data of the quantum
critical point in the transverse  quantum \(XY\) model. The first
column displays the number of significant digits that has been used from
the data to perform the Benford analysis. The quantities in parentheses in
the first row denote the corresponding observables. The numbers in the
right-bottom \(4 \times 4\) box represent the finite-size scaling
exponents, and for example, \(q_{md}(M_z) = 2.10\) for the first
two significant digits implies that if we choose to work with only the first two
significant digits of \(M_z^B\) near the QPT in the transverse \(XY\)
model, the transition point can be detected in a system of finite size
\(N\) with an ``efficiency" \(\lambda_c^N = \lambda_c + \alpha \lambda^{-2.10}\). Here, the word ``efficiency" is used to imply that the
above relation provides a quantitative understanding as to how far we are
from the thermodynamic limit in a system of size \(N\) with respect to the
QPT in the system.
}
\label{tab:table1}
\end{table}
\end{center}
Once the Benford violation parameter is obtained,   for the  the case of transverse magnetzation, we choose a certain number of significant digits, and fit the data to a cubic polynomial 
\begin{equation}
\tilde{a} x^3+\tilde{b} x^2+\tilde{c} x+\tilde{d},
\end{equation}
using the method of least 
squares.
The choice of the power of the polynomial is dictated by the shape of the new data on the field axis.

At $x=-3\tilde{b}/\tilde{a}$, the local maxima of the derivative of the fitted curve is obtained. 
This point gives  the $\lambda_{c}^{N}$ corresponding to the system size  $N$. 
The QPT point, $\lambda_{c}^{N}$, scales with the size of  the system as
\begin{equation}
\lambda_{c}^{N}= \lambda_c+\alpha N^{-q},
\end{equation}
 where $\alpha$ is a constant, for a given number of significant figures.
 and eventually reaches, in the thermodynamic limit, to
 $\lambda_{c}=1$.

Table \ref{tab:table1} shows the values of the finite-size scaling exponents  obtained from the data of Benford violation 
of transverse magnetization and nearest neighbor correlator, $T_{xx}$, for up to the first four significant digits.
It is evident from the table that the exponents increase very little with the 
increase of number of significant digits. 
It therefore follows that 
the very first significant digit possesses sufficient 
information to predict the occurrence of the quantum phase transition in the thermodynamic limit of  the system, and consideration of further digits do not provide a significant advantage in terms of the finite-size scaling exponents. 
This result is potentially  encouraging from the point of view of an experimentalist who can only afford to work with data which is noise-free only to a first few significant digits.

\section{Conclusion} 
\label{sec:conc}
Benford's law is based on  observations that in  data obtained from natural  phenomena and mathematical tables,  the frequencies of occurrences of lower digits as the few leading significant digits is more common as compared to that of the higher ones. In particular, the number ``1" is claimed to appear most often as the first significant digit. This has been adduced by performing analysis on  large data sets obtained from a range of events. Testing Benford's law has usually been concentrated on the   first significant digit. However, recent studies~\cite{Diekmann}  propose that one should also consider digits beyond the first significant one.  In this paper, we study the quantum phase transition in the quantum transverse $XY$ model using Benford's law. 
We use up to  the first four significant digits of data corresponding to transverse magnetization and other physical quantities for detecting the quantum phase transition. Moreover, we perform finite-size scaling analysis in all the instances. 
We find that the measurement of the first significant digit of an observable, in this model,
is enough to detect the quantum phase transition, with a relatively high finite-size scaling exponent. The value of the exponent does not grow significantly by consideration of Benford analysis for a higher number of significant digits. A high finite-size scaling exponent guarantees that the thermodynamic properties of the quantum phase transition can be mimicked by relatively moderate system sizes. This is especially significant given that the physical system under consideration can, with currently available technology, be imitated in moderate-sized laboratory systems, e.g. in cold gases~\cite{cold-atom}.  Note that the Benford analysis up to  four significant digits is possibly sufficient, as beyond that the Beford prediction is close to the uniform distribution. The realization that the first or a first-few significant digits suffice to analyze a certain physical phenomenon is an important one, because it helps us to analyze real physical data, which may be noisy and not reliable after a first few significant digits.  


\section{Acknowledgment}
A.B. acknowledges support of the Department of Science and Technology (DST), Government of India,
through the award of an INSPIRE fellowship. S. S. R. acknowledges the INFOSYS scholarship for senior students. We acknowledge computations performed at the cluster computing facility at
Harish-Chandra Research Institute.

\end{document}